\documentclass[manuscript]{aastex}

\slugcomment{}

\shorttitle{Sgr A$^\ast$ VLBI monitor}
\shortauthors{Tsuboi et al.}
\begin{document}

\title{No Microwave Flare of Sagittarius A$^\ast$ \\around the G2 Periastron Passing}

\author{Masato \textsc{Tsuboi} and Yoshiharu \textsc {Asaki}}
\affil{Institute of Space and Astronautical Science(ISAS), Japan Aerospace Exploration Agency,\\
3-1-1 Yoshinodai, Chuo-ku, Sagamihara, Kanagawa 252-5210, Japan}
\email{tsuboi@vsop.isas.jaxa.jp}
\author{Osamu \textsc{Kameya}  }
\affil{Mizusawa VLBI Observatory, National Astronomical Observatory of Japan, 2-12, Hoshigaoka, Mizusawa, Oshu, Iwate 023-0861, Japan}

\author{Yoshinori \textsc{Yonekura} and Yusuke \textsc{Miyamoto} }
\affil{Center for Astronomy, Ibaraki University, 2-1-1 Bunkyo, Mito, Ibaraki
310-8512, Japan}

\author{Hiroyuki \textsc{Kaneko}, Masumichi \textsc{Seta}, and Naomasa \textsc{Nakai} }
\affil{Division of Physics, Faculty of Pure and Applied Sciences,  University of Tsukuba, Tsukuba, Ibaraki, 305-8571, Japan}

\author{Hiroshi \textsc{Takaba} and Ken-ichi \textsc{Wakamatsu} }
\affil{Faculty of Engineering, Gifu University, 1-1, Yanagito, Gifu, Gifu 501-1193, Japan}

\author{Makoto \textsc{Miyoshi} }
\affil{National Astronomical Observatory of Japan, 2-21-1, Osawa, Mitaka, Tokyo 181-8588, Japan}

\author{Yoshihiro \textsc{Fukuzaki}}
\affil{Geospatial Information Authority of Japan, 1, Kitasato, Tsukuba 305-0811, Japan}

\author{Kenta \textsc{Uehara}}
\affil{Department of Astronomy, the University of Tokyo, Bunkyo, Tokyo 113-0033, Japan}
\and
\author{Mamoru \textsc{Sekido}}
\affil{Kashima Space Research Center, National Institute of Information and Communications Technology(NICT), 893-1 Hirai, Kashima, Ibaraki 314-8501, Japan}

\begin{abstract}
In order to explore any change caused by the G2 cloud approaching, we have monitored the flux density of Sgr A$^\ast$ at 22 GHz from Feb. 2013 to Aug. 2014 with a sub-array of Japanese VLBI Network . The observation period included the expected periastron dates. The number of observation epochs was 283 days. We have observed no significant microwave enhancement of Sgr A$^\ast$ in the whole observation period. The average flux density in the period is $S_\nu=1.23\pm0.33$ Jy. The average is consistent with the usually observed flux density range of Sgr A$^\ast$ at 22 GHz. 
\end{abstract}


\received{October 16, 2014}
\accepted{October 26, 2014}

\section{Introduction}
Sagittarius A$^\ast$ (Sgr A$^\ast$) is a compact source from radio wavelength to X-ray associated with the Galactic center super massive black hole (GCBH), which is located just at the dynamical center of the Galaxy (\cite{Reid}) and has the mass of $M_\mathrm{GC}\sim4\times10^6 M_\odot$ (\cite{Ghez}, \cite{Gillessen2009}). The observed size of Sgr A$^\ast$ is proportional to inverse square of observation frequency $\Delta\theta \mathrm{[mas]}= 1.3\lambda \mathrm{[cm]}^2$  in microwave and millimeter ranges (\cite{Lo}). The relation indicates that the real picture of Sgr A$^\ast$ has been hidden by electron scattering around itself although submillimeter VLBI is now developing quickly to break the barrier (e.g. \cite{Doeleman}). 
 
Recently it was found by IR precision astrometry observations that a small gas cloud is approaching Sgr A$^\ast$ (\cite{Gillessen2012}). The cloud, called ``G2", has the estimated mass of $3M_\mathrm{earth}$. It was predicted that G2 would come closer the periastron distance of $R_\mathrm{peri}\simeq 2000 R_s\simeq 200$ au in the spring of 2014 ($2014.25\pm0.06$; \cite{Gillessen2013}, $2014.21\pm0.14$; \cite{Phifer}). The approaching of the G2 cloud to Sgr A$^\ast$ is a golden opportunity to explore the vicinity of the GCBH using it as a test probe. There were some predictions of increase in emission from Sgr A$^\ast$ induced by the interaction of G2 with the prior existing accretion flow (e.g. \cite{Narayan}, \cite{Sadowski}) or tidal heating of the cloud around the GCBH (e.g.  \cite{Saitoh}).   Although the nature of G2 is still in controversy, G2 is expected to be a gas cloud or a star with dusty envelop (e.g. \cite{Scoville}).  Nevertheless, it is understandable that G2 may give some perturbation to the accretion flow around the GCBH because G2 is seen to be somewhat extended. If any emission enhancement begins, it is very important for study of the mechanism of the event to detect the initial raising phase by ourselves and to alert world-wide community to observe the phenomenon. 
Therefore, we made the project to monitor Sgr A$^\ast$ by VLBI at 22 GHz as frequently as possible. 
Because we don't know  fundamentally when the flux density increase by the G2 cloud approaching will begin, continuous monitor is necessary to detect it. The observation frequency of 22 GHz is expected to be fairly available even in summer of Japan although there is a large year-to-year variation in the weather condition.
A jet eruption from Sgr A$^\ast$ might occur successively. It is also important to observe it continuously even after the event. 

In millimeter and sub-millimeter ranges, the flux density of of Sgr A$^\ast$ is highly variable (e.g. \cite{Miyazaki}, \cite{Mauerhan}). Large flares with several handred percent have been observed frequently.  Although the flux density at 22 GHz of Sgr A$^\ast$ had been monitored for 3 years with VLA, there was no flare with over 100 \% amplitude in the monitor  (\cite{Herrnstein}). If we detect a flare significantly beyond this limit, we can conclude safely that this flare is an unusual phenomenon.

\section{Telescopes and Observations}
We have monitored the flux density of Sgr A$^\ast$ at 22.2 GHz from Feb. 2013 to Aug. 2014 with a sub-array of Japanese VLBI Network (JVN) (\cite{Doi}) in order to explore any change caused by the G2 cloud approaching. The sub-array consists of the Mizusawa 10-m radiotelescope (RT), the Takahagi/Hitachi 32-m RT, and the Gifu 11-m RT, and the observed data were recorded by mean of the K5/VSSP32 VLBI data acquisition (DAQ) system (Kondo et al. 2006).
The Tsukuba 32-m RT and the Kashima 34-m RT have joined it when they met to the observation schedules.  The DAQ system at each RT temporary stored VLBI data with 2 bit sampled 32 MHz bandwidth (128 Mbps data rate) in commercially used hard disk drives. The data was transmitted via Internet, immediately after every observation finishes, to the ISAS where a software cross correlator which has been developed by NICT (Kondo et al. 2004) was running. We could obtain the correlation results within a half day after a daily observation performed. The details of the observation system will be shown in another publication.

The angular size of Sgr A$^\ast$ at 22.2 GHz is expected to be $\Delta\theta= 2$ mas by the relation mentioned above. Because the sub-array has projected baselines of 150- 300 km or the beam size (fringe spacing) is about  13 mas, we can observe the flux density of Sgr A$^\ast$ avoiding the flux density decrease by partially resolved-out. On the other hand, Sgr A$^\ast$ is embedded in the strong extended emission surrounding itself. The high spatial resolution must be required to observe selectively the flux density of Sgr A$^\ast$ itself suppressing the contamination from the extended structure. Therefore, the sub-array of the JVN is most suitable for the daily flux density monitor of Sgr A$^\ast$. 

An induced synchrotron flux with the G2 cloud approaching must be centered  around the bow shock by the G2 gas cloud in the accretion flow of the GCBH (\cite{Sadowski}, \cite{Sadowskib}). 
The angular separation between Sgr A$^\ast$ itself and the possible bright spot is expected to be up to 25 mas around the G2 periastron passing assuming that the distance to the Galactic center  is 8 kpc. The two spots can be detected separately with an additional observation by the sub-array. If a new adjoining spot is detected in the flare, this proves that expected electron acceleration is occurred by the interaction between the G2 gas cloud and the accretion flow.

Typical system noise temperatures of RTs of the sub-array were 100-200 K except summer season. The observing time was 10 minute for each object. 
Three quasars, NRAO530, J1626-2951, and J1924-2914, were observed in a day as flux calibrators in the monitor. Sgr A$^\ast$ was observed twice in a day. 
The flux calibrators are bright quasars and located nearby Sgr A$^\ast$, but certainly variable in the time scale of several months. Unfortunately, all RTs of the sub-array don't have a beam switch system for measurement of the total flux density.  Therefore, their flux densities of the quasars were calibrated by the comparison with NGC7027 using the Nobeyama 45-m telescope in order to guarantee the absolute flux density accuracy during the periastron passing. NGC7027 is a bright young planetary nebula with the flux density which is assumed to be $S_\nu=5.5$ Jy at 22.2 GHz (e.g. \cite{Perley}). 
The procedure to obtain the flux density of Sgr A$^\ast$ is as the following.  All 30-m class RTs of the sub-array are closely located in the same prefecture, Ibaraki prefecture. The baselines between them are not sufficient  to suppress the contamination from the surrounding extended emission around Sgr A$^\ast$. On the other hand, two 10-m class RTs, the Mizusawa 10-m RT and the Gifu 11-m RT, are geographically located  300 km in north and 300 km in west apart from the 30-m class RTs, respectively.  Therefore, correlated amplitudes for the two baselines of the Mizusawa 10-m RT-- a 30-m class RT and the Gifu 11-m RT -- a 30-m class RT were used. They are averaged to improve the SN ratio. The  flux density of Sgr A$^\ast$, $S_\nu \mathrm{(Sgr A^\ast)} $, is given by,
$$
S_\nu \mathrm{(Sgr A^\ast)} =A\mathrm{(Sgr A^\ast)}\times \frac{\sqrt[3]{S_\nu\mathrm{(QSO1)}S_\nu\mathrm{(QSO2)}S_\nu\mathrm{(QSO3)}}}{\sqrt[3]{A\mathrm{(QSO1)}A\mathrm{(QSO2)}A\mathrm{(QSO3)}}},
$$
where $A\mathrm{(QSO)}$  is the averaged correlated amplitude of a quasar.  $S_\nu\mathrm{(QSO)}$ is the  flux density which are calibrated by the method based on NGC7027 using the Nobeyama 45-m telescope as mentioned above. 
Geometric mean is useful for suppressing the effect of unexpected flux density change of the quasars.
The absolute calibration uncertainty of the observation is estimated to be $\sim20$\%, which is fairly larger than the statistical error.

\section{Results}
Fig. 1 shows the light curve of Sgr A$^\ast$ at 22.2 GHz in this monitor (also see \cite{Tsuboi2013a}, \cite{Tsuboi2013b}, \cite{Tsuboi2013c}, \cite{Tsuboi2013d}, \cite{Tsuboi} and \cite{Tsuboib}). The horizontal axis is  the elapsed day (DOY)  from 1 Jan, 2013.  The monitor had been performed from Feb. 25, 2013 to Aug. 12, 2014. The number of the observations in the monitor was 283 days.  The error bars show only statistical errors ($\pm1\sigma$) of each data. First of all, we have observed no strong microwave flare of Sgr A$^\ast$ around the expected G2 periastron epochs which are DOY=$456.25\pm32.85$ day (solid line arrow in Fig.1; Gillessen et al. 2013) and DOY=$441.65\pm51.10$ day (broken line arrow in Fig.1; \cite{Phifer}), respectively.  The average flux densities during  the expected periastron periods are $S_\nu(26)=1.19\pm0.21$ Jy and $S_\nu(48)=1.13\pm0.19$ Jy, respectively. The errors show the standard deviation of the observed flux densities. The number in the parentheses is the number of the observations in the epoch. 
The observed flux densities are consistent with those of the NRAO public data with JVLA (open circles in Fig.1; \cite {Chandler}, \cite {Chandler2} and \cite {Chandler3}, \cite{Sjouwerman}).

It has been reported that the radio flux induced by the bow shock will be reached to the peak seven or nine months earlier than the periastron epoch because the bow shock in the accretion flow precedes the G2 cloud itself on the orbit (\cite{Sadowskib}). According to the prediction, the flux density by the bow shock should increase remarkably during DOY$\sim200-280$ day (dot line arrow in Fig.1). Because this epoch unfortunately corresponds to the summer rainy season in Japan,  there is  a long intermission of the observation DOY$=210-245$ day.  Significant enhancement was not detected in the observed flux densities of Sgr A$^\ast$ in the epochs of DOY$=200-210$ day and DOY$=245-280$ day. In addition, 
there is the NRAO public data at 21.2 GHz of $S_\nu=1.27\pm0.13$ Jy  at DOY$=220$ day. Although flare with short duration cannot be ruled out by these data, the expected significant enhancement with long duration did not occur.

The average flux density in the whole observation period is $S_\nu=1.23\pm0.33$ Jy.
The observed flux densities are consistent with those of the NRAO public data,  $S_\nu=1.10\pm0.17$ Jy.
Meanwhile the average flux density for the 3 years monitor with VLA (\cite{Herrnstein}) is $S_\nu=0.93\pm0.16$ Jy.  Although the average and standard deviation of the measured flux densities in this monitor are practically consistent with the values for the 3 years monitor, our values are slightly larger than the values. 
Sgr A$^\ast$  in our monitor may be relatively active. However, the activity should be within the previously observed range (e.g. \cite{Yusef-Zadeh}). 
We have observed no unusual event of Sgr A$^\ast$ in the whole observation epoch.

 \section{Discussion}
 There is no significant enhancement of Sgr A$^\ast$ at 22.2 GHz during the periods including the expected periastron and the expected intensity peaks. The straigtforward and convicing interpretation of the negative  detection is probably that G2 is a star or a tightly bounded star cluster although it may have any dusty envelope.  The possibility had been discussed by several authors  (e.g. \cite{Phifer}, \cite{Eckart}, \cite{Zaja}, \cite{Scoville}).  This interpretation is also consistent with the surviving image of G2 around the periastron obtained by the latest IR observations (\cite{Witzel} ).  
The destruction condition of the dusty envelope by tidal force of the GCBH is roughly given by,
$$
\frac{R_{\mathrm peri}}{r_{\mathrm G2}}\leqslant\Bigl(\frac{M_{\mathrm GC}}{m_{\mathrm G2}}\Bigr)^{1/3},
$$
where $m_{\mathrm G2}$ is the mass  of G2 and $r_{\mathrm G2}$ is the radius of the dusty  envelope around G2. If G2 is a star, it must have the mass of $m_{\mathrm G2}\gtrsim 1M_\odot$.  The destruction condition is not satisfied for the inner envelope with $r_{\mathrm G2}<$ 1 au; $R_{\mathrm peri}/r_{\mathrm G2}>(M_{\mathrm GC}/m_{\mathrm G2})^{1/3}$ even around the periastron distance, $R_{\mathrm peri}\sim 200$ au, although the condition for the outer envelope with $r_{\mathrm G2}>$ a few au is barely satisfied (also see \cite{Murray-Cla}).  In this case, G2 hardly induce the expected microwave enhancement because it has small cross section. A tightly bounded star cluster also corresponds to such case.  

In the case that G2 is a gas cloud as reported initially, which is still supported by a recent IR observation (\cite{Pfuhl}), there may be several possibilities to explain  it. 
One of them is that  the magnetic field in the accretion flow is too strong to make the bow shock wave in the accretion flow. The bow shock preceding the G2 cloud plays a critical role in the acceleration of synchrotron emitting relativistic electrons (\cite{Sadowski}, \cite{Sadowskib}).  A microwave enhancement is not caused if the bow shock is drowned out  in the accretion flow. When the Alfv\'en velocity around Sgr A$^\ast$, $V_{\mathrm A}$, is faster than the velocity of the G2 cloud,  $V_{\mathrm G2}$, the shock wave is suppressed. $V_{\mathrm A}$ is given by,
$$
V_{\mathrm A}\mathrm{[km s^{-1}]}= \frac{2200H \mathrm{[mG]}}{\sqrt{n_{\mathrm H}\mathrm{[cm^{-3}]}}}.
$$
Because the gas density at the Bondi radius of Sgr A$^\ast$ is estimated to be 
$n_{\mathrm H}(R_{\mathrm B})\sim130$
cm$^{-3}$  
from X-ray observation (\cite{Baganoff}),  the Alfv\'en velocity is $V_{\mathrm A}\mathrm{[km s^{-1}]}\sim 200H\mathrm{[mG]}$.
On the other hand,  $V_{\mathrm G2}$ is estimated by,
$$ 
V_{\mathrm G2}\sim\sqrt{\frac{2GM_\mathrm{GC}}{R}}.
$$
Then the velocity of the G2 cloud is $V_{\mathrm G2}\sim 6000$ km s$^{-1}$ around the periastron distance, which is similar to the Bondi radius of Sgr A$^\ast$, $R_{\mathrm peri}\sim R_{\mathrm B}$.  The condition, $V_{\mathrm G2}\lesssim V_{\mathrm A}$, suggests the lower limit of the magnetic field around the periastron distance, $H\gtrsim 30$ mG. 
 
\acknowledgments

We would like to thank Dr. T. Kondo at the National Institute of Information and Communications Technology for providing us with a software correlator. We also thank National Radio Astronomy Observatory for using the public data of the JVLA monitor. 

{\it Facilities:} \facility{JVN}.

\clearpage
\begin{figure}
\begin{center}
\includegraphics[width=15cm]{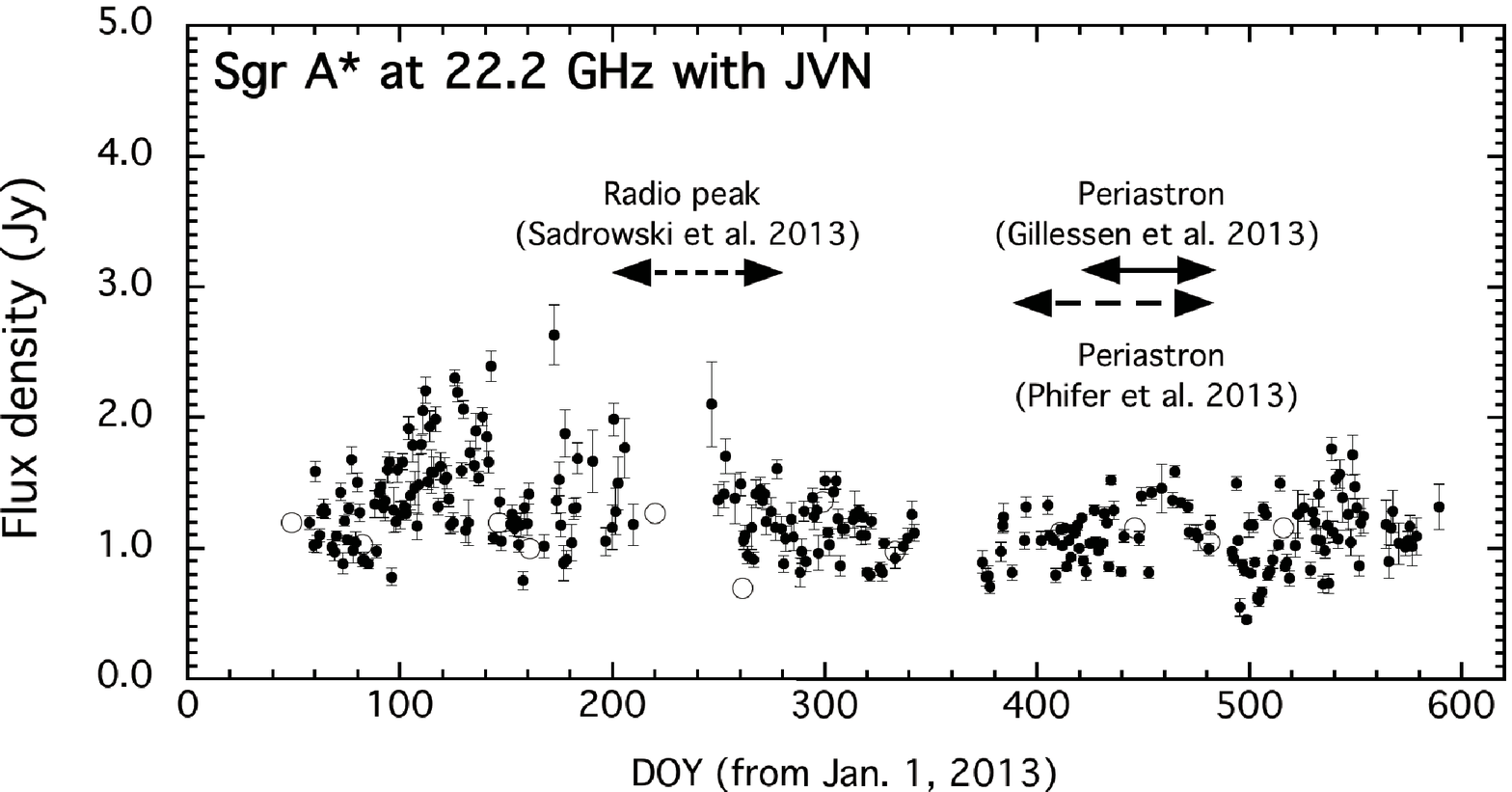}
\caption{The light curve of Sgr A$^\ast$ at 22.2 GHz with the JVN monitor (filled circles). The horizontal axis is  the elapsed day (DOY)  from 1 Jan, 2013. The monitor had been performed from Feb. 25, 2013 to Aug. 12, 2014.  The error bars show only statistical errors ($\pm1\sigma$) of each data. We have observed no significant enhancement of the flux density  of Sgr A$^\ast$ at 22.2 GHz in the whole monitor epoch. Open circles show the flux densities of Sgr A$^\ast$ at 21.2 GHz from the NRAO public data.}
\label{fig1}
\end{center}
\end{figure}

\end{document}